\documentclass[journal]{IEEEtran}
\usepackage{blindtext, graphicx}
\usepackage{listings}
\usepackage{float}
\usepackage{graphicx}
\usepackage[font=small]{caption}
 \usepackage{amsmath}
\usepackage[utf8]{inputenc}
\usepackage{multirow}
\usepackage{multicol}
\usepackage{dblfloatfix}
\usepackage{soul}
\title{Quantification of Tracer Dilution Dynamics: An Exploration into the Mathematical Modeling of Medical Imaging}

\author{
   \IEEEauthorblockN{Ishmael N. Amartey$^1$, Andreas A. Linninger$^2$, Thomas Ventimiglia$^{2,3}$}
   
   \thanks{Submitted on February 29, 2024. This article draws upon research carried out at the Laboratory for Product and Process Design (LPPD), University of Illinois at Chicago, from May 15 to August 31, 2023, in Chicago, USA. \emph{Asterisk indicates cor- responding author}.} % <-this % stops a space

   \thanks{\IEEEauthorrefmark{1}Ishmael N. Amartey, Department of Statistics and Actuarial Science, Northern Illinois University, Dekalb, Illinois, 60115, USA \\(email: iamartey1@niu.edu).}% <-this % stops a space

   \thanks{$^1${Department of Statistics and Actuarial Science, Northern Illinois University, Dekalb, Illinois, USA.
    }}% <-this % stops a space

    \thanks{$^2${Department of Biomedical Engineering, University of Illinois at Chicago, Chicago, Illinois, USA.
    }}% <-this % stops a space

    \thanks{$^3${Department of Mathematical Sciences, Northern Illinois University, Dekalb, Illinois, USA.
    }}% <-this % stops a space

}

\begin{document}

\maketitle
%\thispagestyle{empty}
%\pagestyle{empty}

%%%%%%%%%%%%%%%%%%%%%%%%%%%%%%%%%%%%%%%%%%%%%%%%%%%%%%%%%%%%%%%%%%%%%%%%%%%%%%%%
\begin{abstract}

Convolution and deconvolution are essential techniques in various fields, notably in medical imaging, where they play a crucial role in analyzing dynamic processes such as blood flow. This paper explores the convolution and deconvolution of arterial and microvascular signals for determining impulse and residue functions from in vivo or simulated data and the derivation of the relationship between the residue function and perfusion metrics such as the Cerebral Blood Flow $(CBF)$, Mean Transit Time $(MTT)$ and Transit Time to Heterogeneity $(TTH)$. The paper presents the spectral derivatives as a technique for recovering the impulse response function from the residue function, detailing the computational procedures involved and strategies for mitigating noise effects.

\end{abstract}

\textbf{\emph{Index Terms}: Perfusion metrics, convolution and deconvolution, spectral derivatives, medical imaging, tracer dilution, gamma variate curve}.

%\end{abstract}
%%%%%%%%%%%%%%%%%%%

\section{Introduction}{\label{sec1}}

The deconvolution of signals offer valuable insights into blood flow dynamics in medical imaging, with the potential for enhanced diagnosis and management of various medical conditions. In this study, we delve into the convolution and deconvolution of two signals representing the dynamics of contrast agent concentration within blood vessels and explore the relationships between key perfusion metrics. The study is 
based on the deconvolution-based CT and brain perfusion measurement by Fieselmann \cite{fieselmann2011deconvolution} and the study of spectral methods by Trefethen \cite{trefethen2000spectral}.

Section \ref{sec2} presents the convolution and deconvolution of signals in the Fourier domain and how vital perfusion data could be estimated with the residue and arterial input function (AIF).
The relation of key perfusion metrics and a detailed mathematical approach to deriving the relationship between the residue function, $MTT$, $TTH$, $C_a$, $C_t$, and $h(t)$ is discussed in section \ref{sec3}.

We discussed the spectral derivatives as a technique for recovering the impulse response function $h(t)$ from the residue function $k(t)$ in section \ref{sec4} and the need to extend $k(t)$ to be even for accurate reconstruction.

Section \ref{sec5} and \ref{sec6} presents the conclusions emanating from the study and the direction of future work to enhance the accuracy of reconstructing $k(t)$ and other important perfusion parameters.

%%%%%%%%%%%%%%%%%%%%%%%%%%%%%%%%%%%%%%%%%%%%%%%%%%%%%%%%%%%%%%%%%%%

\section{Deconvolution of two signals}{\label{sec2}}

Blood entering the arterial inlet has different paths to travel and varying transit times to reach a destination. The duration required for reaching a specific voxel can be represented as a random variable conforming to a Gamma distribution at the venous outlet \cite{davenport1983derivation}, hence a deconvolution method can be adopted to arrive at the distribution for the average contrast agent concentration $c_{voi}(t)$ within the volume of interest \cite{fieselmann2011deconvolution}.

\begin{figure}[H]
\centering
\includegraphics[width=8.5cm, height= 4 cm]{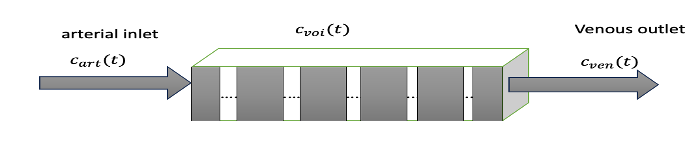}
\caption{Segmented blood vessel}
\label{fig20}
\end{figure}

Fig. \ref{fig20} is a model of interest through which blood flows from a single inlet and is drained from a corresponding single outlet with transit time $t$. Assuming a stationary probability density distribution $h(t)$ of transit times. When a bolus is injected, it enters the vessel through the arterial inlet and is diluted with the blood. The local contrast agent concentrations $c_{art}(t)$ and $c_{ven}(t)$ can be measured directly at the point of entry and exit. The contrast agent concentration $c_{ven}(t)$ at the venous outlet can be computed from the contrast agent concentration $cart(t)$ at the arterial inlet by convolving it with the probability density function $h(t)$ as follows.

We define
\begin{align}
    K(t)\ =\ A\ (\ r(t))
    \label{eq41}
\end{align}
Where $A$ is a constant and $r(t)$ is the residual of CDF of $h(t)$ given as.

\begin{align}
    Residual=r\left(t\right)=1-\int_{0}^{t}h\left(\tau\right)d\tau
    \label{eq42}
\end{align}
for $t\geq 0$ and $r(t)=0$ otherwise.

\begin{align}
    c_{ven}(t)\ =\ \int_{-\infty}^{+\infty}{c_{art}(t-\tau)h(\tau)d\tau}
    \label{eq43}
\end{align}
and $c_{voi}(t)$ which is the average contrast agent concentration in the total volume is given as 
\begin{align}
    c_{voi}(t)\ =CBF(\ \rho_{voi}\ \int_{-\infty}^{+\infty}{c_{art}(\tau)h(t-\tau)d\tau}
    \label{eq44}
\end{align}

\begin{align}
    c_{voi}(t) = CBF(\rho_{voi} \cdot (c_{art} \otimes r))(t)
    \label{eq45}
\end{align}

\begin{align}
    c_{voi}(t)=A\ (\ {(c}_{art}\ \otimes\ K)
    \label{eq46}
\end{align}

where $CBF(\ \rho_{voi} = A$, is a constant representing the multiplication of the Cerebral blood flow (CBF) and is the mean density of the total volume under consideration ($V_{voi})$ respectively. To deconvolve to arrive at $K$ we used the Fourier transform as follows

\begin{align}
    c_{voi}(t)\ =\ c_{art}\ \otimes\ K
    \label{eq47}
\end{align}

\begin{align}
    \mathcal{F}\left(c_{voi}(t)\right)\ \ =\ \mathcal{F}\left(c_{art}(t)\right)\ \ \mathcal{F}\left(K\right)\ \ 
    \label{eq48}
\end{align}

\begin{align}
    \mathcal{F}\left(K\right)\ =\ \frac{\mathcal{F}\left(c_{voi}(t)\right)}{\mathcal{F}\left(c_{art}(t)\right)}
    \label{eq49}
\end{align}

\begin{align}
    K\ =\ \mathcal{F}^{-1}\left(\frac{\mathcal{F}\left(c_{voi}(t)\right)}{\mathcal{F}\left(c_{art}(t)\right)}\right)
    \label{eq50}
\end{align}

The gamma variate curve is normalized to attain a peak value of 1 by the factor

\begin{align}
    \frac{h-min(h)}{max(h)-min(h)}\ 
    \label{eq51}
\end{align}

where $h$ is the gamma variate distribution function.

\subsection{Convolution and Deconvolution Graphs}
\begin{figure}[H]
\centering
\includegraphics[width=8.5cm, height= 4 cm]{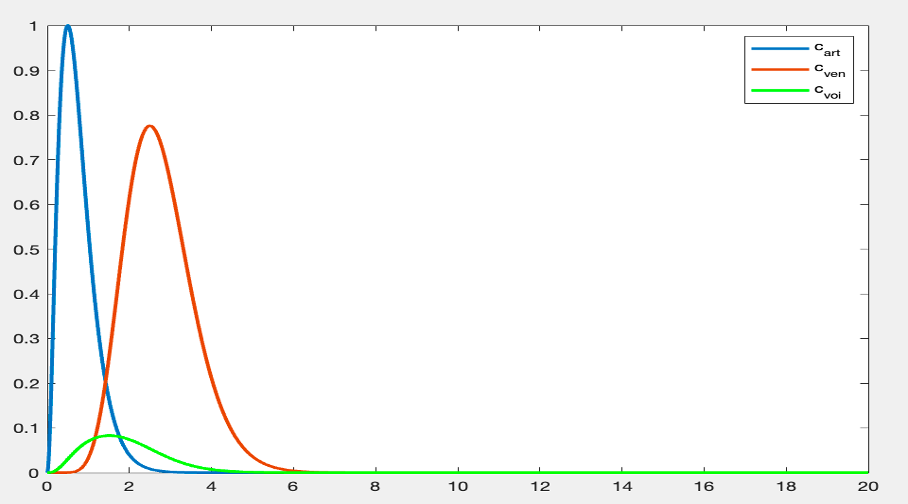}
\caption{Time concentration curves for artificial $c_art(t)$ and $h(t)$ for the construction of $c_{ven}(t)$ and $c_{voi}(t)$}
\label{fig21}
\end{figure}

\begin{figure}[H]
\centering
\includegraphics[width=8.5cm, height= 4 cm]{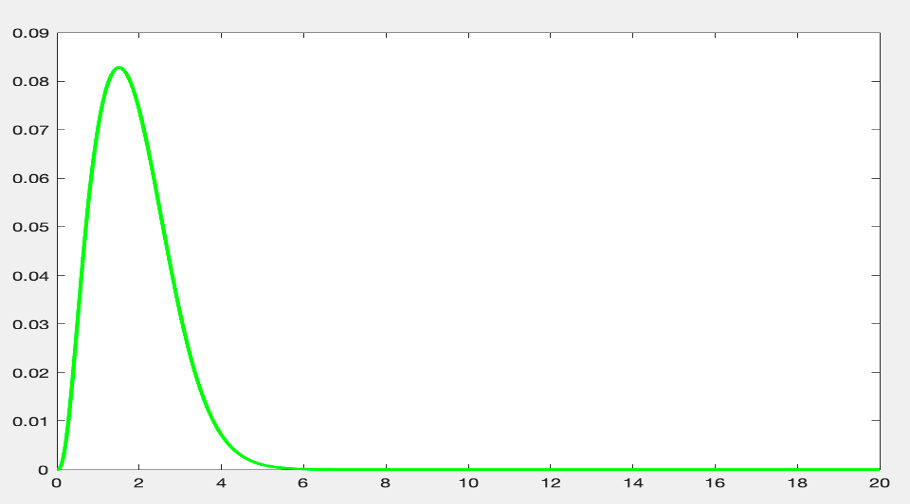}
\caption{A zoomed view of $c_{voi}(t)$}
\label{fig22}
\end{figure}

\begin{figure}[H]
\centering
\includegraphics[width=8.5cm, height= 4 cm]{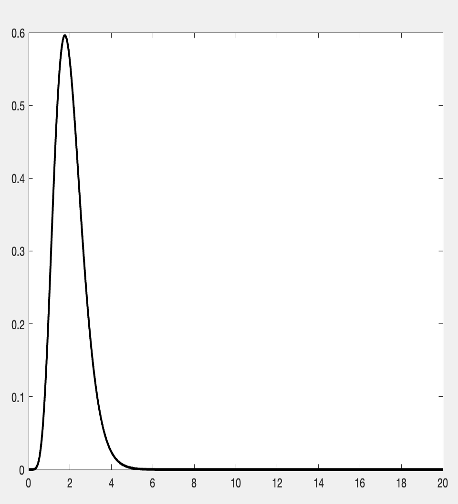}
\caption{Gamma variate curve of  $h(t)$}
\label{fig23}
\end{figure}

\begin{figure}[H]
\centering
\includegraphics[width=8.5cm, height= 4 cm]{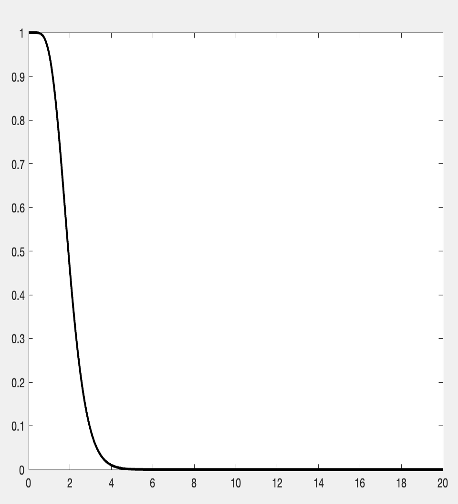}
\caption{Residual curve of $r(t)$}
\label{fig24}
\end{figure}

Fig. \ref{fig21} depicts the $AIF (c_{art})$ and the convolutions in eq. (\ref{eq43}) and eq. (\ref{eq46}). We first construct the $c_{art}$ and the impulse function $h(t)$ to follow a gamma variate function like in eq. \ref{eq52} then the residual function $(K)$ is computed using eq. (\ref{eq41}). The convolution of the $AIF (c_{art})$ and $K$ is $c_{voi}(t)$ which is the tissue signal for the average contrast volume of interest, and the convolution of $c_{art}$ and the impulse function h(t) is the $c_{ven}$. The zoomed plot of $c_{voi}(t)$ is shown in Fig. \ref{fig22} and impulse function $h(t)$ and scaled residual function are shown in Fig. \ref{fig23} and Fig. \ref{fig24} respectively.

\begin{figure}[H]
\centering
\includegraphics[width=8.5cm, height= 4 cm]{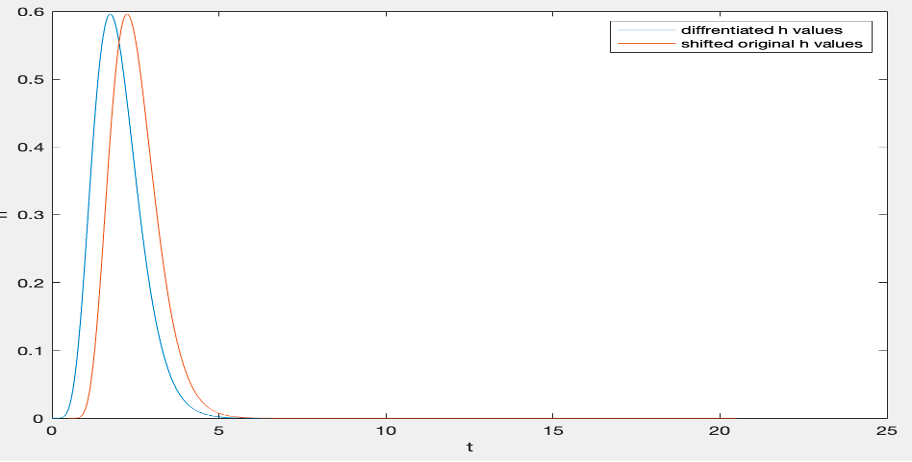}
\caption{Gamma variate plot of the original $h$ values and recovered $h$ values from numerical differentiation}
\label{fig25}
\end{figure}

Given $c_{art}(t)$ and $c_{voi}(t)$ we can reconstruct $K$ using the Fourier Transform in eq. (\ref{eq50}) and reconstruct $h(t)$ by doing a numerical differentiation of $-K/A$. Fig. \ref{fig25} shows the graphs of the original $h(t)$ values and the recovered $h(t)$ values from the numerical differentiation of $-K/A$. 

\section{Relation of key perfusion metrics: $CBF$, $MTT$, $TTH$ to weighted residue function $k(t)$}{\label{sec3}}
Impulse response which conforms to a gamma variate function can be defined as 
\begin{align}
    h(t) = \frac{1}{\Gamma(K)Q^K} t^{K-1} e^{-\frac{t}{Q}}
    \label{52}
\end{align}

where $K>0$ and $Q>0$.
In perfusion analysis, the significance of $K$ and $Q$ lies in their role in determining critical blood perfusion data such as the Mean Transit Time $(MTT)$ and the Transit Time to Heterogeneity $(TTH)$. The first moment of $h(t)$ gives the $MTT$ which is the product of $K$ and $Q$, expressed as $MTT = KQ$, and the second moment of $h(t)$ which is the variance of transit times is the Transit Time Heterogeneity $(TTH)$ expressed as $CTH= KQ^2$. The third and fourth moments can be used to characterize skewness and kurtosis, which may be a useful metric for characterizing abnormal perfusion patterns. The MRI equations are given as 

\begin{align}
    {C}_{a}\ =\ {AIF}
    \label{eq53}
\end{align}
where $AIF$ is the concentration of tracer in the arterial side which is a gamma variate function and

\begin{align}
    C_{t} = CBF \cdot \rho_t (C_a \otimes r)
    \label{eq54}
\end{align}

where $CBF$ is the Cerebral blood flow and $r$ is the residue function.

\subsection{Derivation of eq.\ (\ref{eq54})}

The mean transit time $(MTT)$

\begin{align}
    {MTT}\ =\ \int_{{0}}^{{t}}{{\tau h}({\tau}){d\tau}}
    \label{eq55}
\end{align}

From which we get the residual function in eq. (\ref{eq42}).
The residual function tells the quantity of tracer amount left in the volume of interest at time $t$ and continually decreases over time till it reaches zero. The quantification of accumulated tracers that have entered $(m_{in})$ and left $(m_{out})$ the volume of interest during the time interval $[0, t]$ is expressed as 

\begin{align}
    {m}_{{in}}\ =\ {F}\int_{{0}}^{{t}}{{C}_{a}\left({\tau}\right){d\tau}}
    \label{eq56}
\end{align}

\begin{align}
    {m}_{{out}}\ =\ {F}\int_{{0}}^{{t}}{{C}_{v}({\tau}){d\tau}}
    \label{eq57}
\end{align}

where $F$ is the volume flow assumed to be constant over time. At time $t$ the quantity of tracers in the volume of interest is expressed as 

\begin{align}
    {m}_{t}\ =\ {m}_{{in}}\ - {m}_{{out}}
    \label{eq58}
\end{align}
\begin{align*}
    = F\int_{0}^{t} (C_a(\tau) - C_v(\tau)) \, d\tau
\end{align*}

by convolving ${C}_{a}\left({t}\right)$ and the probability density function $h(t)$, we can produce ${C}_{v}\left({t}\right)$ hence we can express ${C}_{v}\left({t}\right)$ as

\begin{align}
    C_v(t) = \int_{-\infty}^{+\infty} C_a(\xi)h(t-\xi) \, d\xi
    \label{eq59}
\end{align}

since the delta function $\delta(t)$ integrates to one we can convolve it with ${C}_{a}\left({t}\right)$ as follows

\begin{align}
    C_a(t) = \int_{-\infty}^{+\infty} C_a(\xi) \delta(t-\xi) \, d\xi
    \label{eq60}
\end{align}

by substituting eq. (\ref{eq59}) and (\ref{eq60}) into eq. (\ref{eq58}) we get

\begin{align}
    m_t = F\int_{0}^{t} \left(\int_{-\infty}^{+\infty} {C_a(\xi)\delta(\tau-\xi) \, d\xi - C_a(\xi)h(\tau-\xi) \, d\xi} \right) \, d\tau
    \label{eq61}
\end{align}

changing the order of integration yields

\begin{align}
    m_t\ =\ F\int_{-\infty}^{+\infty}{C_{art}(\xi)}\left(\int_{0}^{t}\left(\delta(\tau-\xi)-h(\tau-\xi)\right)d\tau\right)\ d\xi
    \label{eq62}
\end{align}

let $\tau^\prime=\tau-\xi$ then

\begin{align}
    m_t\ =\ F\int_{-\infty}^{+\infty}{C_a(\xi)}\left(\int_{-\xi}^{t-\xi}{\left(\delta(\tau^\prime)-h(\tau^\prime)\right)d\tau^\prime}\right)\ d\xi
    \label{eq63}
\end{align}

notice that $\int_{-{\xi}}^{{t}-\ {\xi}}{\left({\delta}({\tau}^\prime)- {h}({\tau}^\prime)\right) {d}{\tau}^\prime} ={r}({t}-{\xi})$, hence 

\begin{align}
    m_t=\ F\int_{-\infty}^{+\infty}{C_a(\xi)}r(t-\xi)\ d\xi
    \label{eq64}
\end{align}

with the $CBF$ defined as

\begin{align}
    CBF\ =\ \frac{F}{V_t\cdot\rho_t}
    \label{eq65}
\end{align}

where ${C}_{t}$ is the average amount of contrast in the total volume of interest and $\rho_{t}$ is the Mean density of the total volume under consideration $({m}_{t})$, then substituting ${F}= {CBF}\cdot {V}_{t}\cdot \rho _ {t}$ 
into eq. (\ref{eq64}) gives us

\begin{align}
    {m}_{t} &= {CBF} \cdot {V}_{t} \cdot {\rho}_{t} \int_{-\infty}^{+\infty} {C}_{a}(\xi) \, r(t-\xi) \, d\xi
    \label{eq66}
\end{align}

\begin{align*}
    \frac{m_t}{V_t} &= CBF \cdot \rho_t \int_{-\infty}^{+\infty} C_a(\xi) \, r(t-\xi) \, d\xi
\end{align*}

\begin{align}
    C_t=\ CBF\cdot \rho_t\int_{-\infty}^{+\infty}{C_a(\xi)}r(t-\xi)\ d\xi
    \label{eq67}
\end{align}

\begin{align}
    C_t=\ CBF\cdot \rho_t\ {(C}_a\otimes\ r)(t)
    \label{eq68}
\end{align}

so for

\begin{align}
    K(t)\ =\ CBF\cdot \rho_t\cdot\ r(t)
\label{eq69}
\end{align}

\begin{align}
    C_t=\ \left(C_a\otimes K\right)(t)
    \label{eq70}
\end{align}

From eq. (\ref{eq68}), the Fast Fourier Transform $(fft)$ can be estimated as follows

\begin{align}
    {\hat{C}}_t=CBF\cdot \rho_t\left({\hat{C}}_a\otimes\hat{r}\right)
    \label{eq71}
\end{align}

\begin{align}
    CBF(\hat{r})\ =\ \hat{K}\ =\ \frac{{\hat{C}}_t}{{\hat{C}}_a}
    \label{eq72}
\end{align}

\begin{figure}[H]
\centering
\includegraphics[width=18cm, height= 10cm]{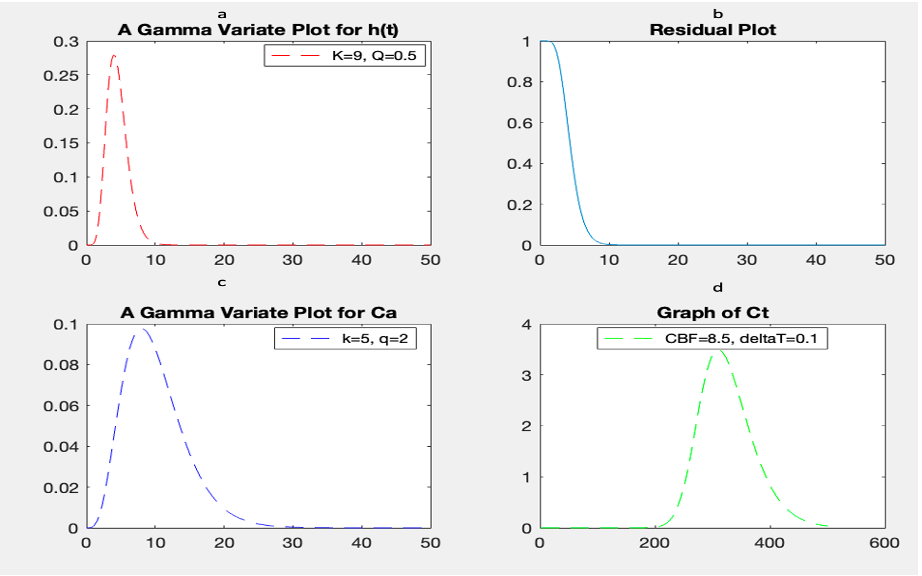}
\caption{Plots of $C_a$, $C_t$, and $R$}
\label{fig26}
\end{figure}

The construction of $h(t)$, $C_t$, and $C_a$ is done by choosing values for $K$ and $Q$. The graphs are shown in Fig. \ref{fig26}.
Fig. \ref{fig26}a is the crude gamma variate graph we began with, with $K-1 (\alpha)$ = 9, and $Q (\beta)$ = 0.5. This is the function we would attempt to reconstruct from the input function of $C_a$ and $C_t$. Fig. \ref{fig26}b is the residue function generated from the $CDF$ of $h(t)$ as stated in eq. (\ref{eq42}). Since $r$ is the residue function from the $CDF$, it attains its maximum at 1 and decays to 0, however since the $CBF$ is a multiplier of $r$, the maximum $K$ value would equal the value of the $CBF$. That is

\begin{align}
    CBF\ =\ max\ \left(\frac{{\hat{C}}_t}{{\hat{C}}_a}\right)\ =\ max(\hat{K})
    \label{eq73}
\end{align}

\begin{align}
    K\ =\ slope\ +1
    \label{eq74}
\end{align}

\begin{align}
    Q\ =\ \frac{t_{max}}{slope}
    \label{eq75}
\end{align}

in Fig. \ref{fig26}c we have an artificial gamma variate curve to represent $C_a$ which was convolved using the Fast Fourier Transform to produce Fig. \ref{fig26}d.

From eq. (\ref{eq72}) we can recover the signals of $k(t)$ using the inverse Fourier transform in a similar manner as eq. (\ref{eq50}) as follows

\begin{align}
    K(t)\ =\ \mathcal{F}^{-1}\left(\frac{{\hat{C}}_t}{{\hat{C}}_a}\right)=IFFT(\ FFT\left(C_a\right)/FFT(C_t))
    \label{eq76}
\end{align}

from which the $CBF$ can be recovered as the maximum value of $K(t)$. To recover $MTT$ and $TTH$ from $K(t)$ which is the first and second moments respectively, we do the following. Since

\begin{align}
    h=-\frac{1}{CBF}\frac{dk}{dt}
    \label{eq77}
\end{align}

using integration by parts

\begin{align}
    \int u dv=uv-\int v du
    \label{eq78}
\end{align}

let

\begin{align}
    u = k(t) \quad \Rightarrow \quad du = \frac{dk}{dt} \cdot dt
    \label{eq79}
\end{align}

\begin{align*}
    dv = dt \quad \Rightarrow \quad v = t
\end{align*}

thus, \begin{align}
      \int k\left(t\right) dt=tk\left(t\right)-\int t\frac{dk}{dt} \ dt
    \label{eq80}
\end{align}

substitute the expression for $h=-\frac{1}{CBF}\frac{dk}{dt}$  into eq. (\ref{eq80}) to get

\begin{align}
    \int k\left(t\right) dt=tk\left(t\right)-\int t\left(-CBFh\right) \ dt
    \label{eq81}
\end{align}

\begin{align*}
    \int k\left(t\right) dt=tk\left(t\right)+CBF\int th \ dt
\end{align*}

Now, integrating both sides over the interval $[0,\infty)$:

\begin{align}
    \int_{0}^{\infty} k(t)\, dt = \lim_{A \rightarrow \infty} \left(Ak(A) + CBF \int_{0}^{A} th\, dt\right)
    \label{eq82}
\end{align}

As $A$ approaches infinity, $k\left(A\right)$ would approach a constant if the function is well-behaved. The integral $\int_{0}^{A}th\ dt$ would also approach a finite value. Thus, as $A$ tends to infinity, we get:

\begin{align}
    \int_{0}^{\infty} k(t)\, dt = CBF \int_{0}^{\infty} th\, dt
\label{eq83}
\end{align}

\begin{align*}
    \frac{1}{CBF} \int_{0}^{\infty} k(t)\, dt = \int_{0}^{\infty} th\, dt
\end{align*}

\begin{align}
    MTT = \frac{1}{CBF} \int_{0}^{\infty} k(t)\, dt
\label{eq84}
\end{align} 

Similarly, $k\left(t\right)$ relates to $TTH$ as

\begin{align}
    TTH = \frac{2}{CBF} \int_{0}^{\infty} t \cdot k(t)\, dt - MTT^2
\label{eq85}
\end{align} \\

\begin{figure}[h]
\centering
\includegraphics[width=8.5cm, height= 6cm]{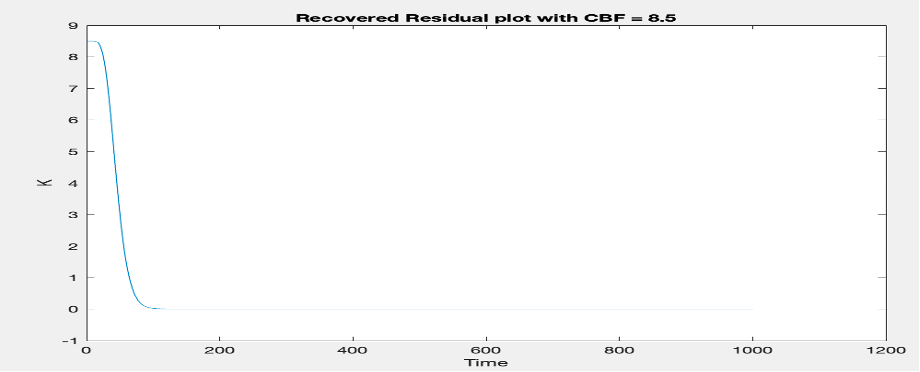}
\caption{Recovered residual plot from the input function  $C_a$ and $C_t$}
\label{fig27}
\end{figure}

%\twocolumn
\section{Spectral Derivatives}{\label{sec4}}
To recover $h(t)$ from $k(t)$ we use the spectral derivative. 
Suppose we have a signal $y_j$ for $j=1,2,\ldots,n$. The $FFT$ gives, for $k=1,\ldots,\ n$
\\ \\ \\ \\ \\ \\ \\ \\ \\ \\ \\ \\ \\ \\ \\ \\ \\ \\ \\ \\ \\ \\ \\ 
\\ \\ \\ \\ \\ \\ \\ \\ \\ \\
\begin{align}
    {\hat{y}}_k=\sum_{j=1}^{n}{y_j\exp{(-2\pi i(j-1)(k-1)/n})}
    \label{eq86}
\end{align}

$y_j$ can be recovered by

\begin{align}
    y_j=\frac{1}{n}\sum_{k=1}^{n}{{\hat{y}}_k\exp({2\pi i\left(j-1\right)\left(k-1\right)}/n)}
    \label{eq87}
\end{align}

Let $x_j=\left(j-1\right)h=\left(j-1\right)P/n$ to define a range of $x$ values on the interval $\left[0,P\right]$ for which $y\left(x_j\right)=y_j$. Then,

\begin{align*}
    x_j=\left(j-1\right)P/n
\end{align*}

\begin{align}
    \frac{x_j}{p}=\frac{\left(j-1\right)}{n}
    \label{eq88}
\end{align}

substituting eq. (\ref{eq88}) into eq. (\ref{eq87}) gives us

\begin{align}
    y_j=\frac{1}{n}\sum_{k=1}^{n}{{\hat{y}}_k\exp ({2\pi i x_j\left(k-1\right)}/P)}
    \label{eq89}
\end{align}

replacing $x_j$ with any real valued $x\in[0,P]$ yields an interpolating function $p(x)$ that satisfies $p\left(x_j\right)=y_j$.

\begin{align}
    \left(x\right)=\frac{1}{n}\sum_{k=1}^{n}{{\hat{y}}_k\exp ({2\pi ix\left(k-1\right)}/P)}
    \label{eq90}
\end{align}

we then assign the value of the numerical derivative at each sample point $x_j$ to equal $p\prime(x)$. 

\begin{align}
    p^\prime\left(x\right)=\frac{1}{n}\sum_{k=1}^{n}{{\hat{y}}_k\frac{2\pi i(k-1)}{P}\ \exp ({2\pi ix\left(k-1\right)}/P)}
    \label{eq91}
\end{align}

The algorithm for computing the spectral derivative is thus,
\begin{enumerate}
    \item Compute ${\hat{y}}_k=FFT(y_j)$
    \item Form the vector of coefficients ${{\hat{p}}_k=\hat{y}}_k\frac{2\pi i(k-1)}{P}$
    \item 	Obtain the vector of spectral derivatives at the sample points: $y_j^\prime=IFFT({\hat{p}}_k)$. 
\end{enumerate}

\begin{figure}[H]
\centering
\includegraphics[width=8.5cm, height= 5cm]{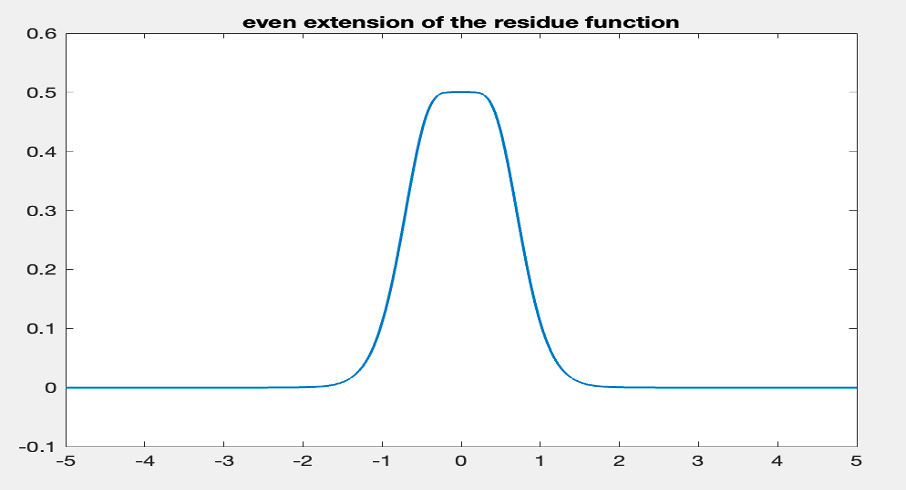}
\caption{Even extension of $k(t)$}
\label{fig28}
\end{figure}

\begin{figure}[H]
\centering
\includegraphics[width=8.5cm, height= 6.2cm]{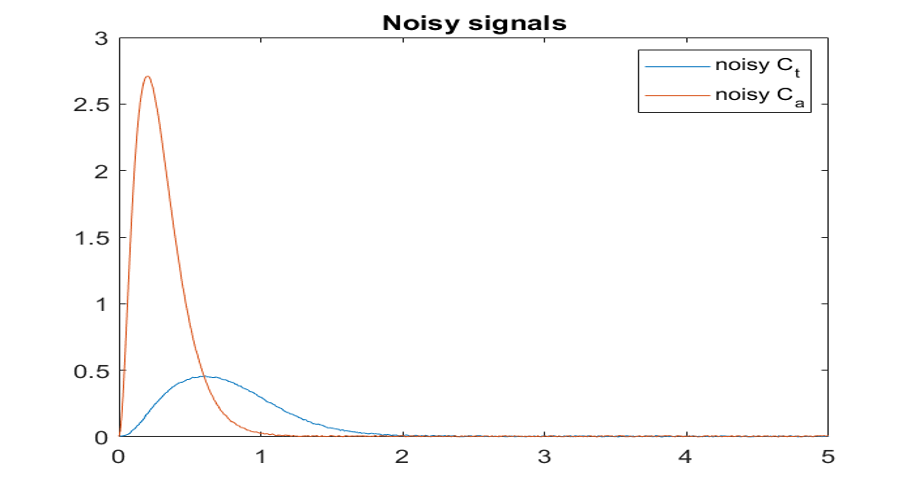}
\caption{Recovered noisy signals of $C_t$ and $C_a$}
\label{fig29}
\end{figure}

\begin{figure}[H]
\centering
\includegraphics[width=8.5cm, height= 6.2cm]{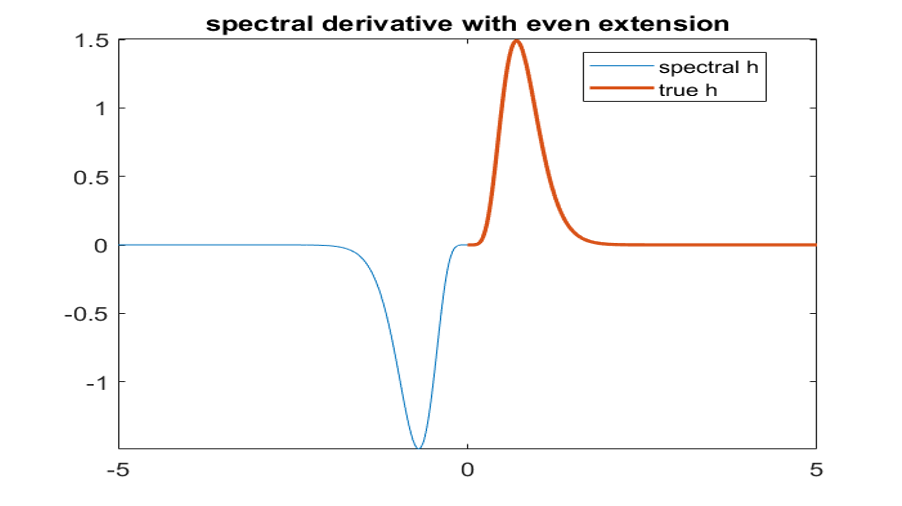}
\caption{Spectral derivative of $h$ showing a mirrored image of original $h$}
\label{fig30}
\end{figure}

The caveat for the spectral derivative requires a function to be periodic without discontinuities but Since $k(t)$ is not periodic it would be extended to be even so it can take the shape of a periodic function to satisfy that condition. Fig. \ref{fig28} shows the extension of $k(t)$ into an even function. By extending $k(t)$ as even, the reconstruction of $h(t)$ is achieved through the spectral derivative perfectly after discarding the negative time components . Fig. \ref{fig29} shows the successful retrieval of the noisy $C_a$ and $C_t$ via the spectral derivative procedure. As can be seen in Fig. \ref{fig30}, the spectral derivative after the even extension recovers the original $h(t)$ but in a mirrored manner, extending the $k(t)$ to be even would lead to the exact reconstruction of $h(t)$ by just cutting off the negative time components. It is important to also acknowledge that reconstructing $K$ with the spectral derivative is very sensitive to noise therefore any slight addition of noise to $C_a$ and $C_t$ produces a highly noisy $k(t)$ which in turn would not provide accurate vital perfusion $C_a$ and $C_t$ estimates, Fig. \ref{fig32} depicts such  $k(t)$.

\begin{figure}[H]
\centering
\includegraphics[width=8.5cm, height= 6.2cm]{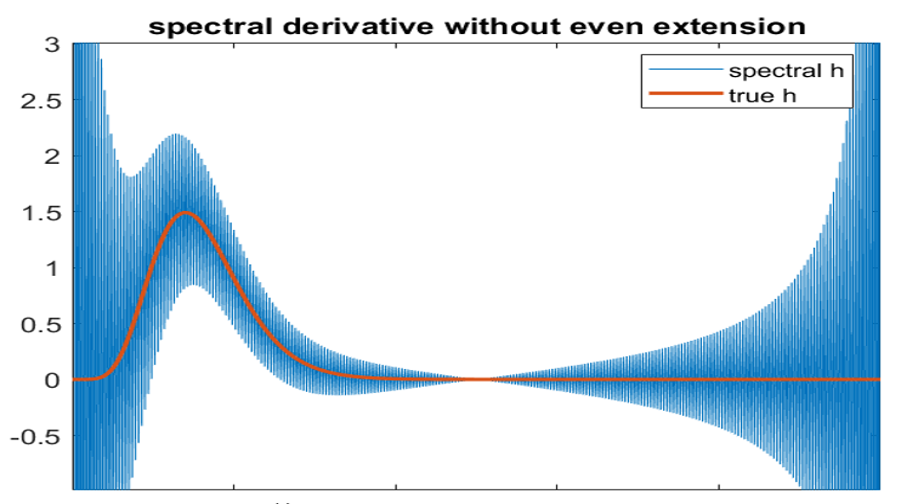}
\caption{Comparison of prescribed $h\left(t\right)$ with the reconstruction from the spectral derivative. Without extending $k(t)$ to be even, the derivative is overwhelmed by erroneous modes.}
\label{fig31}
\end{figure}

\begin{figure}[H]
\centering
\includegraphics[width=8.5cm, height= 6cm]{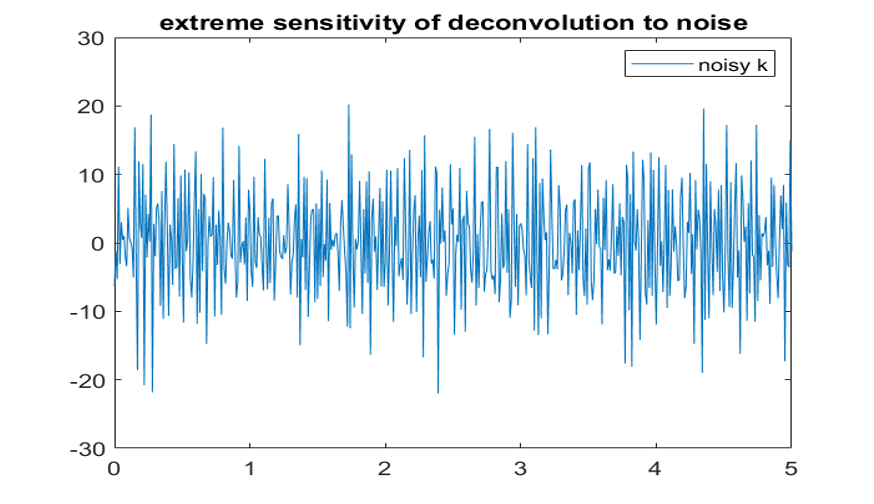}
\caption{The addition of any noise whatsoever wrecks the reconstruction of $k(t)$}
\label{fig32}
\end{figure}

\begin{figure}[H]
\centering
\includegraphics[width=8.5cm, height= 5cm]{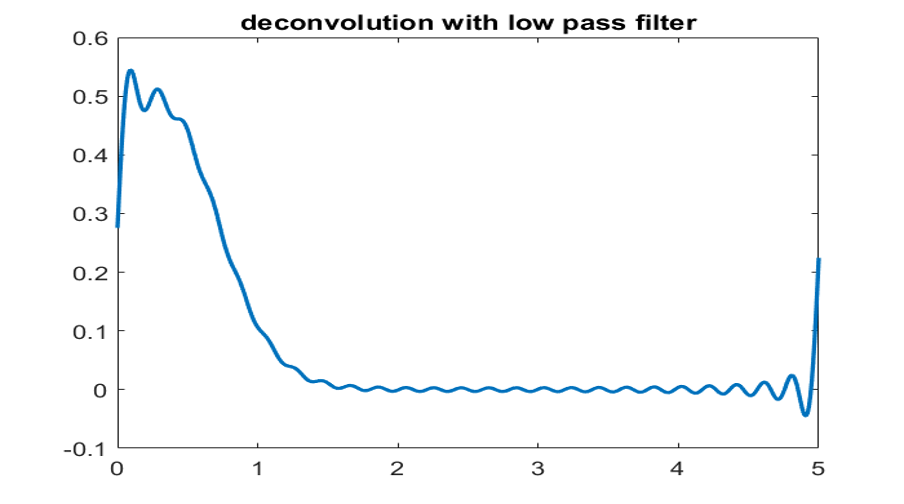}
\caption{Reconstruction of the residual function $k(t)$ with low pass filter. The estimation of $CDF$ as $max{k(t)}$ is significantly affected by Gibbs phenomenon, which causes greater oscillations at the endpoints of the domain.}
\label{fig33}
\end{figure}

\begin{figure}[H]
\centering
\includegraphics[width=8.5cm, height= 6cm]{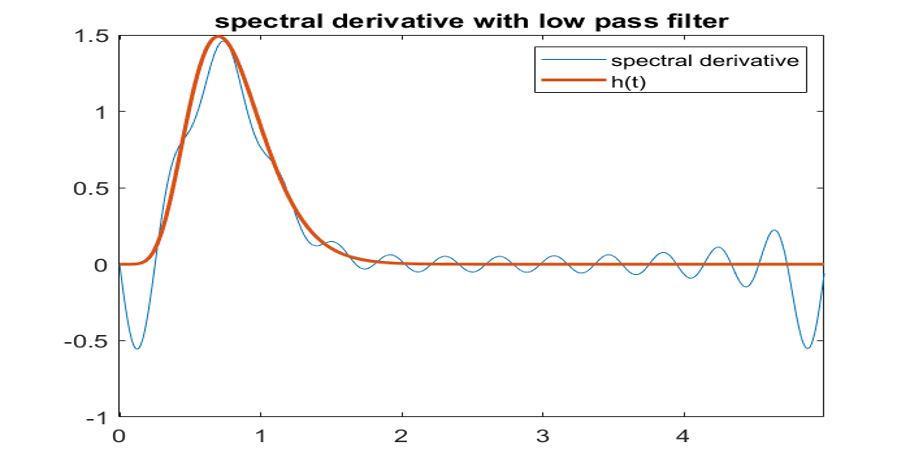}
\caption{The spectral derivative with high frequencies removed reconstructs the transfer function $h(t)$. The estimate for $h(t)$ can be filtered more by fitting it to a Gamma variate function, as in the Madsen procedure}
\label{fig34}
\end{figure}

\section{Conclusion}{\label{sec5}}

The estimation of the residue function, $r(t)$, and impulse response functions, $h(t)$, can be achieved through the deconvolution of tracer intensity curves. The residue function signifies the number of tracers retained in the volume of interest while $h(t)$ is modeled as a Direc delta function \cite{amartey2024derivation}. Deconvolving perfusion signals requires an arterial input function stemming from either a venous outlet to estimate the impulse function, $h(t)$, or from an aggregate tissue average to estimate the residue function, $r(t)$. To successfully deconvolve signals, the use of the Discrete Fourier Transforms (DFT) must be adopted to enable the numerical differentiation of the residue function to arrive at $h(t)$ but the DFT requires the residue function to be periodic.

\section{Future work}{\label{sec6}}

Deconvolution is a vital part of medical imaging analysis, particularly in estimating key perfusion parameters. This paper has shed light on how to recover the residue function $r(t)$ and impulse function $h(t)$ from tracer curves via deconvolution methods, however the process when characterized does not properly recover the original curve hence the need to explore other methods that are robust to noise.

While this study has provided a foundational understanding of deconvolution techniques in medical imaging, there remains ample room for further research and development particularly to minimize error and improve the accuracy of perfusion metrics.

Overall, the application of deconvolution techniques holds great promise for enhancing the capabilities of medical imaging systems, thus, the focus on future work should be robust non-parametric ways to estimating key perfusion metrics with less error and assumptions.

%%%%%%%%%%%%%%%%%%%%%%%%%%%%%%%%%%%%%%%%%%%%%%%%%%%%%%%%%%%%%%%%%%%%
\section*{Acknowledgment}
I would like to thank Dr. Andreas Linninger, the director of the Laboratory for Product and Process Design (LPPD) and Thomas Ventimiglia for their support throughout the course of this study.


%%%%%%%%%%%%%%%%%%%%%%%%%%%%%%%%%%%%%%%%%%%%%%%%%%%%%%%%%%%%%%%%%%%%%%%%%%%%%%%%
\nocite*
\bibliographystyle{IEEEtran}  
\bibliography{main}

\begin{thebibliography}{1}
\providecommand{\url}[1]{#1}
\csname url@samestyle\endcsname
\providecommand{\newblock}{\relax}
\providecommand{\bibinfo}[2]{#2}
\providecommand{\BIBentrySTDinterwordspacing}{\spaceskip=0pt\relax}
\providecommand{\BIBentryALTinterwordstretchfactor}{4}
\providecommand{\BIBentryALTinterwordspacing}{\spaceskip=\fontdimen2\font plus
\BIBentryALTinterwordstretchfactor\fontdimen3\font minus \fontdimen4\font\relax}
\providecommand{\BIBforeignlanguage}[2]{{%
\expandafter\ifx\csname l@#1\endcsname\relax
\typeout{** WARNING: IEEEtran.bst: No hyphenation pattern has been}%
\typeout{** loaded for the language `#1'. Using the pattern for}%
\typeout{** the default language instead.}%
\else
\language=\csname l@#1\endcsname
\fi
#2}}
\providecommand{\BIBdecl}{\relax}
\BIBdecl

\bibitem{fieselmann2011deconvolution}
A.~Fieselmann, M.~Kowarschik, A.~Ganguly, J.~Hornegger, and R.~Fahrig, ``Deconvolution-based ct and mr brain perfusion measurement: Theoretical model revisited and practical implementation details,'' \emph{Int J Biomed Imaging}, vol. 2011, p. 467563, 2011.

\bibitem{trefethen2000spectral}
L.~N. Trefethen, \emph{Spectral methods in MATLAB}.\hskip 1em plus 0.5em minus 0.4em\relax Society for Industrial and Applied Mathematics, 2000.

\bibitem{davenport1983derivation}
R.~Davenport, ``The derivation of the gamma-variate relationship for tracer dilution curves,'' \emph{J Nucl Med}, vol.~24, no.~10, pp. 945--948, 1983.

\bibitem{amartey2024derivation}
I.~N. Amartey, A.~A. Linninger, and T.~Ventimiglia, ``The derivation and reconstruction of the gamma variate function for tracer dilution curves,'' 2024.

\end{thebibliography}

\end{document}